\documentstyle[twocolumn,prl,aps,epsfig]{revtex}

\begin{document}

\wideabs{

\title{Excitation of a Dipole Topological Mode in a Strongly Coupled
Two-Component Bose-Einstein Condensate}

\author{J. Williams, R. Walser, J. Cooper, E.A. Cornell, M. Holland}

\address{JILA, National Institute for Standards and Technology and
University of Colorado, Boulder, CO 80309-0440}

\date{Submitted to PRA on April 28, 1999}

\maketitle

\begin{abstract}
Two internal hyperfine states of a Bose-Einstein condensate in a
dilute magnetically trapped gas of ${}^{87}$Rb atoms are strongly
coupled by an external field that drives Rabi oscillations between the
internal states. Due to their different magnetic moments and the force
of gravity, the trapping potentials for the two states are offset
along the vertical axis, so that the dynamics of the internal and
external degrees of freedom are inseparable. The rapid cycling between
internal atomic states in the displaced traps results in an adiabatic
transfer of population from the condensate ground state to its first
antisymmetric topological mode. This has a pronounced effect on the
internal Rabi oscillations, modulating the fringe visibility in a
manner reminiscent of collapses and revivals. We present a detailed
theoretical description based on zero-temperature mean-field theory.

\vspace{.2cm}
PACS numbers(s): 03.75.Fi, 05.30.Jp, 32.80.Pj, 74.50.+r
\end{abstract}
}
\vspace{.2cm}

\section{Introduction} \label{intro}
An intriguing aspect of Bose-Einstein condensation (BEC) in a dilute
atomic gas is that the internal atomic state of the condensate can be
manipulated to produce quite novel systems. A number of interesting
experiments have produced uncoupled multi-component condensates, in
which two or more internal states of the condensate exist together in
a magnetic or optical trap
\cite{Wieman1,Ketterle1,Ketterle2,exp1,exp2}. These experimental
studies, along with their theoretical counterparts, have investigated
various topics such as the ground state of the system
\cite{ho1,esry2,pu2,ho2,ohberg2,ao}, the elementary excitations
\cite{meystre1,busch1,walls3,pu1,esry1,ohberg3,savage}, and the
nonlinear dynamics of component separation \cite{sinatra}.

Many fascinating properties can be studied by applying an external
electromagnetic field that coherently couples the internal atomic
states of the
condensate~\cite{exp3,eschmann,williams1,Villain,ohberg,exp5,blakie,ZhangWalls,ZhangLiu}. In
the experiment described in \cite{exp3}, the relative phase between
two hyperfine components was measured using a technique based on
Ramsey's method of separated oscillating fields \cite{Ramsey} and
these results have motivated further theoretical investigation
\cite{sinatra,eschmann}.  In this system, several key parameters can
be varied over a wide range of values, such as the coupling-field
intensity and frequency, the confining potentials, the total number of
atoms, and the temperature, making this a very rich system to explore.

Several theoretical papers \cite{williams1,Villain,ohberg} have
investigated the weak coupling limit of this system, where the
intensity of the external driving field is very low but is turned on
for a time long compared to the period of oscillation in the magnetic
trap. A clear analogy exists between this system and the Josephson
junction \cite{Barone}. In the Josephson junction, identical particles
in spatially separated condensates are coupled via the tunneling
mechanism
\cite{Walls2,Juha,Walls1,Wright,Legget1,Legget2,Smerzi1,Smerzi2,Clark}.
In this two-component system, however, two distinct internal states of
the condensate are coupled by an applied field. By adjusting the
magnetic fields that confine the atoms, the degree of spatial overlap
between the two components can be controlled.

Recently, there have been both experimental \cite{exp5} and
theoretical \cite{blakie} studies on the dressed states of a driven
two-component condensate, drawing an analogy to the dressed states of
a driven two-level atom in quantum optics. Due to the interplay
between the internal and external degrees of freedom, the condensate
dressed states have spatial structure that depends on the trap
parameters, the mean-field interaction, and the frequency and
intensity of the driving field \cite{blakie}. In the experiment
reported in \cite{exp5}, the dressed states were created via an
adiabatic passage by sweeping the detuning.

In this paper, we focus on the limit of a very strong and sustained
coupling between hyperfine states, which is the situation achieved
experimentally in \cite{exp5}. In that experiment, a BEC of about
$8\times 10^5$ atoms was produced in the $|F=1,M_F=-1\rangle$
hyperfine state of ${}^{87}$Rb, at a temperature close to zero,
$T\approx 0$. The atoms were confined in a time-averaged, orbiting
potential (TOP) magnetic trap by a harmonic potential with axial
symmetry along the vertical axis. An external field was then applied
that coupled the $|1,-1\rangle$ state to the $|2,1\rangle$ hyperfine
state via a two-photon transition. The Rabi frequency was five to ten
times larger than the vertical trap frequency and the detuning could
be adjusted arbitrarily.  Due to their different magnetic moments and
the force of gravity, the two hyperfine states sit in shifted traps
offset along the vertical axis \cite{exp4}. The degree of separation
could be controlled by adjusting the magnetic trapping fields.

The subsequent behavior of the system described in \cite{exp5} was
quite unexpected: after the coupling field was turned on, the Rabi
oscillations between the states appeared to collapse and revive on a
time scale which was long compared to the Rabi period of 3 ms. An
example of this behavior taken from \cite{exp5} is shown in Fig.~1. It
was observed that the period of this modulation increased with
decreasing detuning. The behavior of the system also depended
critically on the separation between the traps for each state.  As the
separation was taken to zero, the effect went away. The evolution of
the density of each component was also very interesting. Each
component cycled between a density profile with one peak and a profile
with two peaks. The two-peaked structure was most clearly visible
around the collapse time, which is $t \approx 20$ ms for the case
shown in Figure 1.
\begin{figure}
  \centerline{\epsfig{file=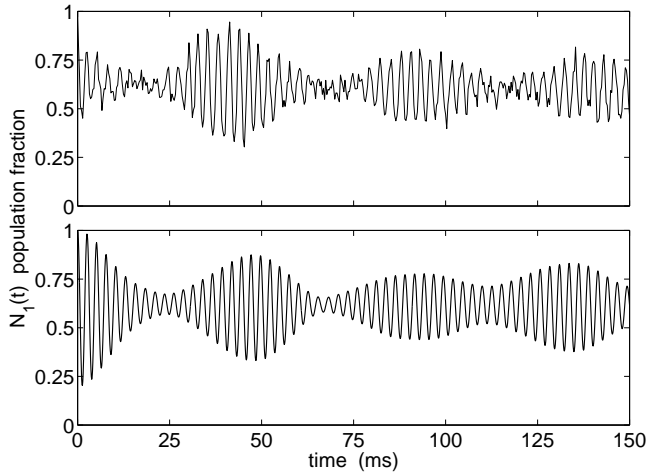,height=2.5in}}
\vspace{0.05cm}
\caption{This plot shows the modulation of the fractional population
in the (1,-1) state. The top line is experimental data
\protect\cite{exp5} while the bottom line is the result of a
numerical calculation of the three-dimensional, two-component
Gross-Pitaevskii equation Eq.~(\ref{main1}) (below). The coupling
strength and detuning were chosen for the calculation to be $\Omega =
350$ Hz and $\delta = -188$ Hz, respectively.}
\end{figure}

A numerical calculation of the three-dimensional, coupled
Gross-Pitaevskii (GP) equations Eq.~(\ref{main1}) (below) describing
the system in the zero temperature limit agrees, at least
qualitatively, with the outcome of the experiment, as shown in Figure
1. Of course, the agreement between the numerical integration of
Eq.~(\ref{main1}) and the laboratory data does not in itself provide
an intuitive explanation of the underlying physical mechanism
responsible for the collapse-revival behavior. In this paper, we
present a detailed analysis of this problem, and arrive at a rather
simple model that explains the major features of the system's
behavior.

Before presenting the details of our analysis, it is useful to first
give an overview of the results. There are two main concepts that play
key roles in obtaining an intuitive understanding of this problem.
First, there is a clear separation of time scales: the period of Rabi
oscillations between internal states is much shorter than the period
of the trap.  That is, the internal dynamics occur on a much shorter
time scale than the motional dynamics of the system.  It is therefore
useful to go to a frame rotating at the effective Rabi frequency. In
this rotating frame, we show that there exists a weak coupling between
the low lying motional states which is proportional to the offset
between the two traps. This weak coupling has the effect of modulating
the amplitude of the fast Rabi oscillations in the lab frame.

The second key point is to understand exactly which motional states
are excited.  They are not the linear response collective excitations
(normal modes) that have been studied frequently in the BEC literature
\cite{griffin,stringari3,fetter}. Instead, they are many-particle
topological modes determined by the self-consistent solutions to the
two-component GP equations Eq.~(\ref{selfcons}). The well known vortex
mode~\cite{edwards1,stringari1,sinha,rokhsar1} is one example of such
an excitation in which phase continuity requires quantized circulation
around a vortex core. The related excitation which plays a key role in
this paper does not exhibit circulation but has a node in the
wavefunction amplitude and exhibits odd-parity behavior characteristic
of such a dipole mode. For a single component in the limit of the
uniform gas, the exact solution of this mode is known as a dark
soliton~\cite{Clark}. In that case the scale of the density
perturbation around the node is the healing length and is determined
by a balance of kinetic and mean-field interaction energies. In the
problem we consider here, however, it is necessary to account for the
mean field of the the remaining population in the condensate ground
state. Consequently the two modes---the ground state and the dipole
mode---are inextricably linked and must be determined self
consistently.

We first present a detailed theoretical analysis in
Section~\ref{theory}.  After making several reasonable approximations,
we arrive in Section~\ref{twomode} at the two-mode model -- a
simplified description that encapsulates the essential properties of
the system. In Section~\ref{casestudy} we present results of numerical
calculations that illustrate the behavior of the system and we compare
our model with the exact numerical solution of the coupled GP
equations. We finally summarize our work in Section~\ref{conclusion}
and suggest further studies based on our understanding of this
phenomenon.

\section{Theoretical Description} \label{theory}

The following theoretical development was motivated by the experiment
described in \cite{exp5}. Therefore, we have not tried to keep our
calculations general, but instead have made several assumptions based
on that particular situation.  However, our approach could easily be
extended to treat a broader class of systems.  We give a brief
discussion in the conclusion of the paper about possible extensions of
this work to other interesting systems.

We begin this section by writing down the coupled mean-field
equations, valid for zero temperature, that describe this driven,
two-component BEC. In Section~\ref{extint} we rewrite the mean-field
equation in a direct-product representation that clearly separates out
the external and internal dynamics. We then go to a frame rotating at
the effective Rabi frequency in Section~\ref{rotframe} in order to
focus on the slower motional dynamics of the system.  After making
some approximations in Section~\ref{approx}, we finally arrive at the
main result of our study in Section~\ref{twomode}, the two-mode model.

\subsection{Coupled Mean-field Equations} \label{coupledmfeq}

A mean-field description of this many-body system that includes the
atom-field interaction has been developed, which generalizes the
standard Gross-Pitaevskii (GP) equation to treat systems with internal
state coupling~\cite{williams1,Villain,cirac1}. The resulting
time-dependent GP equation describing the driven, two-component
condensate is
\begin{equation}
\begin{array}{ccc}
i   \left( \! \! \begin{array}{c}  
       \dot{\psi}_1  \\ \dot{\psi}_2  \end{array} \!\! \right)
   \!\!\!  &=& \!\!\! \left( \!\! \begin{array}{cc}  
    H_1^0 + H_1^{\rm{MF}} + 
       \delta/2 & \Omega/2 \\
       \Omega/2 &  H_2^0 + H_2^{\rm{MF}} - \delta/2 
\end{array} \! \right) \!\!   \left( \!\! \begin{array}{c}  
       \psi_1  \\  \psi_2  \end{array} \!\! \right) \,\, .
\end{array}
\label{main1}
\end{equation}
The Hamiltonians describe the evolution in the trap $H_i^0$ and the
mean field interaction $H_i^{\rm{MF}}$ for each component
\begin{eqnarray}
H_i^0 &=& -{1\over{2}} \nabla^2 + {1\over{2}}[({\rho\over\alpha})^2 +
(z + \gamma_i \,z_0)^2] \nonumber \\ H_i^{\rm{MF}} &=& N(\lambda_{ii}
|\psi_i|^2 + \lambda_{ij} |\psi_j|^2) \,\,,
\label{Hs}
\end{eqnarray}
where $\gamma_1=-1$ and $\gamma_2=1$, and $z_0$ is the shift of each
trap from the origin along the vertical axis. The factor
$\alpha=\omega_z/\omega_{\rho}$ is the ratio of axial and radial trap
frequencies. In the experiment reported in \cite{exp5}, $\alpha >
1$. The mean-field strength is characterized by $\lambda_{ij} = 4\pi
a_{ij}/z_{\rm{sho}}$, which depends on the scattering length $a_{ij}$
of the collision.  In general there will be three different values,
one for each type of collision in this two-component gas: $a_{11}$,
$a_{22}$, $a_{12}$.  The detuning between the driving field and the
hyperfine transition is given by $\delta$ while $\Omega$ denotes the
strength of the coupling.  We work in dimensionless units: time is in
units of $1 / \omega_z$, energy is in terms of the trap level spacing
$\hbar \, \omega_z$, and position is in units of the harmonic
oscillator length $z_{\rm{sho}} = \sqrt{\hbar / m_{\rm{Rb}} \,
\omega_z}$. The complex functions $\psi_i({\bf{r}},t)$ are the
mean-field amplitudes of each component, where $i=\{1,2\}$. They obey
the normalization condition $\int (|\psi_i|^2 + |\psi_j|^2) d^3x =
1$. The total population is $N$.

The coupled mean-field equations Eq.~(\ref{main1}) can be solved
numerically using a finite-difference Crank-Nicholson algorithm
\cite{holland1}.  We show results of such calculations in Fig.~1 for
the three-dimensional solution and in Section~\ref{casestudy} for a
one-dimensional version. However, in order to gain a more intuitive
understanding of the behavior shown in Fig. 1, we formulate a
simplified description of the system in the following section.

\subsection{External $\otimes$ Internal Representation} \label{extint}

The coupled mean-field equations Eq.~(\ref{main1}) can be rewritten in
a more illuminating form by making a clear separation of the external
and internal degrees of freedom.  The system exists in a direct-product
Hilbert space ${\mathcal H = H_{\rm{ex}}\otimes H_{\rm{in}}}$, where
$\mathcal H_{\rm{ex}}$ is the infinite-dimensional Hilbert space
describing the motional state of the system in the trap and $\mathcal
H_{\rm{in}}$ is the two-dimensional Hilbert space describing the spin
of the system. A general operator in $\mathcal H$ can be written as a
sum over the direct-product of operators from $\mathcal H_{\rm{ex}}$
and $\mathcal H_{\rm{in}}$. We rewrite Eq.~(\ref{main1}) in this
representation as
\begin{equation}
  i {\partial\over\partial t} |\psi(t)\rangle = [\hat{H}_{0} \otimes \hat{1} + 
  \hat{1} \otimes ({\Omega\over2} 
  \, \hat{\sigma}_x + {\delta \, \over 2} \hat{\sigma}_z) + 
  \hat{H}_{z} \otimes \hat{\sigma}_z] |\psi(t)\rangle
\label{main2}
\end{equation}
where $\{\hat{1},\hat{\sigma}_x,\hat{\sigma}_y,\hat{\sigma}_z\}$ are
the standard Pauli spin matrices. The state of the system
$|\psi(t)\rangle$ in general has a nonzero projection on the internal
states $|1\rangle$ and $|2\rangle$, represented by
$\psi_i({\bf{r}},t)=\langle {\bf{r}} | \langle i |\psi(t)\rangle$,
where $i = \{1,2\}$. The position representations of $\hat{H}_{0}$ and
$\hat{H}_{z}$ are local, i.e. $\langle
{\bf{r}}|\hat{H}_{0}|{\bf{r}}'\rangle = H_{0}({\bf{r}}) \,
\delta({\bf{r}}-{\bf{r}}')$ and $\langle
{\bf{r}}|\hat{H}_{z}|{\bf{r}}'\rangle = H_{z}({\bf{r}}) \,
\delta({\bf{r}}-{\bf{r}}')$, where $H_{0}({\bf{r}})$ and
$H_{z}({\bf{r}})$ are given by
\begin{eqnarray}
 H_{0}({\bf{r}}) &=& -{1\over{2}} \nabla^2 + {1\over{2}}[({\rho\over\alpha})^2 +z^2] + \langle \psi(t)|\hat{P}_{\bf{r}} \otimes \hat{\lambda}_{+} |\psi(t)\rangle\,\, , \nonumber \\
 H_{z}({\bf{r}}) &=& - z_0 \, z + \langle \psi(t)|\hat{P}_{\bf{r}} \otimes \hat{\lambda}_{-} |\psi(t)\rangle \,\, .
\label{HSA}
\end{eqnarray}
The operator $\hat{P}_{\bf{r}}$ is the projector onto the position
eigenstates $\hat{P}_{\bf{r}}=|\bf{r}\rangle \langle \bf{r} |$, and
the matrix representations of $\hat{\lambda}_{+}$ and
$\hat{\lambda}_{-}$ are given as
\begin{eqnarray}
      \hat{\lambda}_{+}  &=& {N \over 2}
         \left( \begin{array}{cc}  
        \lambda_1 + \lambda_{12} & 0 \\ 
        0 & \lambda_2 + \lambda_{12}
\end{array} \right) \,\, , \nonumber
\end{eqnarray}
\begin{eqnarray}
      \hat{\lambda}_{-} &=& {N \over 2}
\left( \begin{array}{cc}  
        \lambda_1 - \lambda_{12}  & 0 \\ 
        0 & \lambda_{12} - \lambda_2
\end{array} \right) \,\, .
\label{lambdaSA}
\end{eqnarray}
Note that the harmonic potential in $\hat{H}_0$ is centered at the
origin.  The mean-field interaction has been rewritten in terms of a
part that acts identically on both components $\langle
\psi|\hat{P}_{\bf{r}} \otimes \hat{\lambda}_{+} |\psi\rangle \otimes
\hat{1}$, and a part that acts with the opposite sign on each state
$\langle \psi|\hat{P}_{\bf{r}} \otimes \hat{\lambda}_{-} |\psi\rangle
\otimes \hat{\sigma}_z$.

The first two terms in Eq.~(\ref{main2}) separately describe the
external and internal dynamics of the system, respectively. The third
term in Eq.~(\ref{main2}), however, couples the internal state
evolution to the condensate dynamics in the trap and can lead to
interesting behavior. If the term $\hat{H}_{z}$ were identically zero,
then the problem would be completely separable in terms of the
external and internal degrees-of-freedom. The term $\hat{H}_{z}$ would
be zero if the trap separation $z_0$ were zero and if the scattering
lengths were all exactly equal. In fact, for ${}^{87}$Rb the three
scattering lengths are nearly degenerate, so the main effect of
$\hat{H}_{z}$ comes from the term $-z_0 z$, which is the difference in
the shifted traps. It causes there to be a spatially varying detuning
across the condensate.

\subsection{Rotating Frame} \label{rotframe}

As previously stated, we are concentrating on the situation where the
coupling strength is large, so that the frequency of the Rabi
oscillations $\Omega$ is significantly larger than the trap frequency
$\nu_z$. In this case, the internal spin dynamics and the motion of
the condensate in the trap occur on two different time
scales. Therefore, it is useful to go to a rotating frame that
eliminates the second term in Eq.~(\ref{main2}) describing the fast
Rabi oscillations between the two internal states. In the rotating
frame, we will be able to understand more clearly how the third term
in Eq.~(\ref{main2}), which couples the motional and spin dynamics of
the condensate, effects the system on a time scale much longer than
the period of Rabi oscillation.

We go to the rotating frame, or interaction picture, by making a
unitary transformation using the operator
\begin{equation}
U_I(t) = e^{-i \, \hat{1}\otimes \, ({\Omega \over2} \hat{\sigma}_x + {\delta 
\over 2} \hat{\sigma}_z) \, t}
\,\, .
\label{U1}
\end{equation}
This can be rewritten in the equivalent form
\begin{equation}
U_I(t) = \hat{1}\otimes(\cos(\Omega_{\rm{eff}} / 2 \,t) \hat{1} - {i
\over \Omega_{\rm{eff}}} \sin(\Omega_{\rm{eff}}/2 \, t)[\Omega
\hat{\sigma}_x + \delta \hat{\sigma}_z]) \,\, ,
\label{U2}
\end{equation}
where $\Omega_{\rm{eff}}=\sqrt{\Omega^2 + \delta^2}$.  The state
vector $|\psi^{(I)}(t) \rangle$ in the rotating frame is related to
the state vector in the lab frame $|\psi(t) \rangle$ by
\begin{equation}
|\psi^{(I)}(t) \rangle = U^\dagger_I \, |\psi(t) \rangle \,\, .
\end{equation}
In the rotating frame, the system evolves according to
\begin{equation}
i {\partial \over \partial t}|\psi^{(I)}(t) \rangle = \hat{H}^{(I)}(t) 
|\psi^{(I)}(t) \rangle \,\, ,
\label{intpic}
\end{equation}
where $\hat{H}^{(I)}(t)$ is the interaction Hamiltonian
\begin{equation}
\hat{H}^{(I)}(t)= \hat{H}_{0} \otimes \hat{1} + \hat{H}_{z} \otimes
(\alpha_{\rm{x}}(t) \hat{\sigma}_{\rm{x}} + \alpha_{\rm{y}}(t)
\hat{\sigma}_{\rm{y}} + \alpha_{\rm{z}}(t) \hat{\sigma}_{\rm{z}}) \,\,
.
\label{HofI}
\end{equation}
Note that $\hat{H}_{0}$ and $\hat{H}_{z}$ are unaffected by the
unitary transformation to the rotating frame.  The time-varying
coefficients $\alpha_{\rm{x}}(t)$, $\alpha_{\rm{y}}(t)$, and
$\alpha_{\rm{z}}(t)$ are
\begin{eqnarray}
  \alpha_{\rm{x}}(t) &=& {\Omega \over \Omega_{\rm{eff}}} {\delta
  \over \Omega_{\rm{eff}}} [1 - \cos(\Omega_{\rm{eff}}t)] \nonumber \\
  \alpha_{\rm{y}}(t) &=& {\Omega \over \Omega_{\rm{eff}}}
   \sin(\Omega_{\rm{eff}}t) \nonumber \\
  \alpha_{\rm{z}}(t) &=& {\delta^2 \over \Omega_{\rm{eff}}^2}
  +  {\Omega^2 \over \Omega_{\rm{eff}}^2} \cos(\Omega_{\rm{eff}}t) \,\,.
\label{alphas}
\end{eqnarray}

\subsection{Approximations} \label{approx}

We now make three simplifications in order to extract out the dominant
behavior of the system. We first note that the coefficients
$\alpha_i(t)$ given in Eq.~(\ref{alphas}) oscillate rapidly at the
Rabi frequency. We expect the system in the rotating frame
$|\psi^{(I)}(t) \rangle$ to evolve on a much slower time scale than
the period of Rabi oscillation. We can utilize this fact in order to
simplify the interaction Hamiltonian $\hat{H}^{(I)}(t)$ given in
Eq.~(\ref{HofI}) by taking the average values of the coefficients
$\alpha_i(t)$ -- this is equivalent to coarse graining
Eq~(\ref{intpic}). The coefficients in Eq.~(\ref{alphas}) become time
independent and reduce to: $\alpha_{\rm{x}}=\delta \, \Omega /
\Omega_{\rm{eff}}^2$, $\alpha_{\rm{y}}=0$, and
$\alpha_{\rm{z}}=\delta^2 / \Omega_{\rm{eff}}^2$.

We make the additional assumption that the system is being driven
close to resonance, so that $\delta/ \Omega_{\rm{eff}} << 1$. We
therefore set $\alpha_{\rm{z}}=0$ since $\alpha_{\rm{z}}$ depends
quadratically on this small parameter. 

Finally, we take advantage of the fact that the scattering lengths for
${}^{87}$Rb are nearly degenerate, with the ratios between inter- and
intra-species scattering lengths given by $\{a_2:a_{12}:a_1\} =\{
0.97:1:1.03\}$ \cite{exp2}. This allows us to simplify the two mean-field
terms appearing in Eq.~(\ref{HSA}). We first make the approximation
$\hat{\lambda}_{+} \approx \lambda\, N \cdot \hat{1}$ by assuming
equal scattering lengths, or $\lambda_1 \approx \lambda_2 \approx
\lambda_{12} \rightarrow \lambda$. We can also simplify the other term
$\hat{\lambda}_{-}$ by assuming that its predominant effect is to
shift the levels slightly. Instead of neglecting it altogether, we
simply replace it by a mean-field shift of the levels $\langle
\psi|\hat{P}_{\bf{r}} \otimes \hat{\lambda}_{-} |\psi\rangle
\rightarrow \delta_{\rm{MF}}$, where the shift is given by
$\delta_{\rm{MF}}= \Delta\lambda\int n^2({\bf{r}},0) d^3x / N$. Here
$n({\bf{r}},0)$ is the total density at $t=0$ and $\Delta\lambda=
(\lambda_1-\lambda_{12})=(\lambda_{12}-\lambda_2)$.  We can absorb it
into the detuning $\delta$ by defining an effective detuning $\delta'$
that includes this mean-field shift $\delta \rightarrow \delta'=
\delta + \delta_{\rm{MF}}$.

After making the above approximations, we can now write the
interaction Hamiltonian $\hat{H}^{(I)}$ from Eq.~(\ref{HofI}) in a much
simpler form
\begin{equation}
\hat{H}^{(I)} = \hat{H}_{0}' \otimes \hat{1} + \hat{H}_{z}' \otimes
\hat{\sigma}_{\rm{x}} \,\, ,
\label{HofI2}
\end{equation}
where the position representations of $\hat{H}_{0}'$ and
$\hat{H}_{z}'$ are local, i.e. $\langle
{\bf{r}}|\hat{H}_{0}'|{\bf{r}}'\rangle = H_{0}'({\bf{r}}) \,
\delta({\bf{r}}-{\bf{r}}')$ and $\langle
{\bf{r}}|\hat{H}_{z}'|{\bf{r}}'\rangle = H_{z}'({\bf{r}}) \,
\delta({\bf{r}}-{\bf{r}}')$, where $H_{0}'({\bf{r}})$ and
$H_{z}'({\bf{r}})$ are given by
\begin{eqnarray}
 \hat{H}_{0}'({\bf{r}}) &=& -{1\over{2}} \nabla^2 + {1\over{2}}
 [({\rho\over\alpha})^2 +z^2] + \lambda \, n({\bf{r}},t) \,\, ,
 \nonumber \\ \hat{H}_{z}'({\bf{r}}) &=& - \beta \, z \,\, .
\label{HSA2}
\end{eqnarray}
The total density is $n({\bf{r}},t)=N \,\langle
\psi(t)|\hat{P}_{\bf{r}} \otimes \hat{1}|\psi(t)\rangle$, and $\beta =
z_0 \, \delta \, \Omega / \Omega_{\rm{eff}}^2$.  For the typical
values of the parameters in the experiment, $\Omega = 400$ Hz,
$\delta' = 100$ Hz, $z_0/z_{\rm{sho}} = 0.1$, this factor is rather
small $\beta \approx 0.02$ in harmonic oscillator units.  Note that
$\hat{H}_{0}'$ still varies slowly in time through the nonlinear
mean-field term, which depends on the density. We refer to this result
Eq.~(\ref{HofI2}) as the coarse-grained, small detuning (CGSD) model to
distinguish it from the two-mode model presented below, which makes
further assumptions.

We have managed to greatly simplify the description of the system by
going to the rotating frame. The first term in Eq.~(\ref{HofI2})
contains the kinetic energy, a harmonic potential centered at the
origin, and a mean-field interaction term depending on the slowly
varying density $n({\bf{r}},t)$.  The second term in Eq.~(\ref{HofI2})
represents a very weak coupling between the two internal states
$|1\rangle$ and $|2\rangle$, and between motional states
$|\phi_n\rangle$ and $|\phi_m\rangle$ via the dipole operator $z$. The
states $|\phi_n\rangle$ and $|\phi_m\rangle$ are the instantaneous
self-consistent eigenmodes of $\hat{H}_{0}'$. In the next subsection
we present a model that assumes only two motional states are coupled,
the self-consistent ground state $|\phi_0\rangle$ and the
self-consistent first-excited state $|\phi_1\rangle$, which has
odd-parity along the z-axis.

\subsection{Two-mode model}\label{twomode}
It is useful to define a basis of motional states with which to
describe the system in the rotating frame. A natural choice is the set
of instantaneous eigenstates of $\hat{H}_{0}'$, which satisfy
\begin{eqnarray}
( -{1\over{2}} \nabla^2 + {1\over{2}} [({\rho\over\alpha})^2 +z^2] +
 \lambda \, n({\bf{r}},t)) \phi_i({\bf{r}}) &=&
 \epsilon_i \, \phi_i({\bf{r}}) \nonumber \\ \int_{-\infty}^{\infty}
 \phi_i({\bf{r}}) \, \phi_j({\bf{r}})\, d^3r &=& \delta_{i,j} \,\, ,
\label{selfcons}
\end{eqnarray}
where the index $i$ refers to all of the relevant quantum numbers that
uniquely specify each eigenstate, $i=\{n_z,n_{\rho},n_{\phi}\}$, given
the cylindrical symmetry of the system. In general, many  modes
can be occupied and the state vector is written
\begin{equation}
  |\psi^{(I)}(t)\rangle = \sum_i[c_i(t)|\phi_i\rangle|1\rangle
  + d_i(t)|\phi_i\rangle |2\rangle] \,\, ,
\label{state_selfcons}
\end{equation}
where $\phi_i({\bf{r}})=\langle {\bf{r}} | \phi_i \rangle$. The
density appearing in Eq.~(\ref{selfcons}) is then
\begin{equation}
  n({\bf{r}},t) = N(| \sum_i c_i(t)\phi_i({\bf{r}})|^2
                  +  | \sum_i d_i(t)\phi_i({\bf{r}})|^2) \,\, .
\end{equation}
It is clear that the set of coupled eigenvalue equations given in
Eq.~(\ref{selfcons}) is nonlinear and requires a numerical procedure
that will converge upon the solution in a self-consistent manner.  The
eigenstates $\phi_i({\bf{r}})$ and eigenenergies $\epsilon_i$ depend
on time implicitly through the coefficients $c_i(t)$ and
$d_i(t)$, however we do not show this time dependence in order to
simplify the notation. We assume that the eigenbasis evolves slowly in
time so that the adiabatic condition is satisfied \cite{Messiah}.

Based on the experiment reported in \cite{exp5} the initial motional state
of the system is $\psi_1^{(\rm{I})}({\bf{r}},0) = \phi_0({\bf{r}}-z_0
{\hat{z}})$; the system is in the ground state of $\hat{H}_{0}'$, but
displaced from the origin along the vertical axis by $z_0$. This
displacement is small compared to the width $w_z$ of the condensate
$z_0/w_z \approx 0.01$. We therefore approximate the initial state of
the system as $|\psi^{(I)}(t)\rangle =|\phi_0\rangle |1\rangle$.

The system in the rotating frame evolves according to the Hamiltonian
described by Eq.~(\ref{HofI2}) and Eq.~(\ref{HSA2}). The term
$\hat{H}_{z}' \otimes \hat{\sigma}_{\rm{x}}$ couples the internal
states $|1\rangle$ and $|2\rangle$ via ${\hat{\sigma}}_{\rm{x}}$.  It
also drives transitions between motional states via the dipole
operator ${\hat{z}}$.  The dipole matrix element $\langle z
\rangle_{ij} = \langle \phi_i | {\hat{z}} | \phi_j \rangle$ is the
largest between neighboring states and falls off quickly as $|i-j|$
increases.  For a small coupling parameter $\beta$, we expect the
coupling to the first excited state $|\phi_1\rangle$ to dominate the
other transitions, making the evolution of the system predominantly a
two state evolution. We therefore make the approximation that the
system occupies only two modes
\begin{equation}
 |\psi^{(I)}(t)\rangle = c_0(t)\,|\phi_0\rangle |1\rangle
 + d_1(t)\,|\phi_1\rangle |2\rangle \,\, ,
\label{twostate}
\end{equation}
where $|\phi_0\rangle$ is the ground state $i=\{0,0,0\}$ and
$|\phi_1\rangle$ is the first excited state with odd parity along the
z-axis $i=\{1,0,0\}$.

If we substitute this ansatz into Eq.~(\ref{intpic}), using the
Hamiltonian described by Eq.~(\ref{HofI2}) and Eq.~(\ref{HSA2}), we
get the equation of motion for the coefficients $c_0(t)$ and
$d_1(t)$
\begin{equation}
\begin{array}{ccc}
i   \left( \! \! \begin{array}{c}  
       \dot{c}_0  \\ \dot{d}_1  \end{array} \!\! \right)
   \!\!\!  &=& \!\!\! \left( \!\! \begin{array}{cc}  
    \epsilon_0  & -\beta \, \langle z \rangle_{01} \\
       -\beta \, \langle z \rangle_{01} &  \epsilon_1 
\end{array} \! \right) \!\!   \left( \!\! \begin{array}{c}  
       c_0  \\  d_1  \end{array} \!\! \right)
\end{array} \,\, ,
\label{twostateevolve}
\end{equation}
where we have neglected the time rate-of-change of the slowly varying
adiabatic eigenbasis. This coupled pair of equations must be solved
numerically by updating the energies $\epsilon_i$ and the dipole
matrix element $\langle z \rangle_{01}$ from solving
Eq.~(\ref{selfcons}) at each time step. However, in order to see how
the behavior depends on the various physical parameters, one can
obtain a simple estimate of the solution by fixing $\epsilon_i$ and
$\langle z \rangle_{01}$ to their initial values.  In this case the
solution of Eq.~(\ref{twostateevolve}) is trivial and is given by
$c_0(t) = \cos(\Omega_{01}/2 \, t) - i (\Delta \epsilon_{01} /
\Omega_{01}) \sin(\Omega_{01}/2 \, t)$ and $d_1(t) = -i (2\beta \langle
z \rangle / \Omega_{01}) \sin(\Omega_{01}/2 \, t)$, where $\Delta
\epsilon_{01} = \epsilon_1-\epsilon_0$ and $\Omega_{01} =
\sqrt{4\beta^2 \langle z \rangle^2 + \Delta \epsilon_{01}^2}$.  In the
rotating frame, the system oscillates between the two states at a
frequency of $\Omega_{01}$, which is much slower than the effective
Rabi frequency $\Omega_{\rm{eff}}$.

The oscillation frequency $\Omega_{01}$ increases with increasing
detuning $\delta'$ and increasing trap separation $z_0$ through the
coupling parameter $\beta$. The amplitude of oscillation depends on
the energy spacing between modes $\Delta \epsilon_{01}$. Based on
numerical calculations, we have found that this effect is enhanced by
the mean-field interaction because $\Delta \epsilon_{01}$ decreases
with increasing population $N$. Also, the dipole matrix element
$\langle z \rangle$ increases with increasing $N$, since the width of
the condensate increases with increasing population.

The solution in the lab frame can be obtained by applying
$U_{\rm{I}}(t)$ from Eq~(\ref{U2}) to $|\psi^{\rm{(I)}}\rangle$ in
Eq.~(\ref{twostate}) to yield
\begin{eqnarray}
 |\psi(t)\rangle &=& (\alpha_1(t) c_0(t)\,|\phi_0\rangle
 + \alpha_2(t) d_1(t)\,|\phi_1\rangle) |1\rangle \nonumber \\
&+& (\alpha_2(t) c_0(t)\,|\phi_0\rangle
 + \alpha_1^*(t) d_1(t)\,|\phi_1\rangle) |2\rangle \,\, , 
\label{twostatelab}
\end{eqnarray}
where $\alpha_1(t) = \cos(\Omega_{\rm{eff}}/2 \,t) - i(\delta' /
\Omega_{\rm{eff}}) \sin(\Omega_{\rm{eff}}/2 \,t)$ and $\alpha_2(t) =
-i (\Omega / \Omega_{\rm{eff}}) \sin(\Omega_{\rm{eff}}/2 \,t)$.
Eq.~(\ref{twostatelab}) is the main result of our study, with which we
can explain the essential properties of the system. During the first
few Rabi cycles $t \approx 1/\Omega_{\rm{eff}}$, the coefficient
$d_1(t) \approx 0$, so that the solution for short times is
$|\psi(t)\rangle = (\alpha_1(t)|1\rangle + \alpha_2(t)|2\rangle) \,
|\phi_0\rangle$. That is, for short times, the internal and external
degrees of freedom appear to be decoupled and the system simply
oscillates rapidly between internal states.  However, for longer
times, the coefficient $d_1(t)$ grows in magnitude as $c_0(t)$
correspondingly decreases.  This results in a modulation of the Rabi
oscillations. Furthermore, a two-peaked structure in the density
appears, associated with the first-excited state $|\phi_1 \rangle$.

\section{Results}\label{casestudy}
The main goal of this section is to illustrate the behavior of the
system by showing results of numerical calculations.  For this
purpose, it is useful to treat the system in only one
dimension---along the vertical axis \cite{williams1}. We also assume
equal scattering lengths throughout this section, so that
$\delta_{\rm{MF}}=0$. Values of most of the physical parameters are
given in Table 1. Values of the remaining parameters are stated for
each case considered in the text.

\subsection{Understanding the dual dynamics}
In Figure 2 we plot the fractional population of state $|1\rangle$,
given by $N_1(t) = \int |\langle z | \langle 1|\psi(t)\rangle|^2 dz$,
for the case of $\Omega = 700$ Hz and $\delta = 100$ Hz. This is a
numerical solution of Eq.~(\ref{main1}). The population is cycling
rapidly at the effective Rabi frequency $\Omega_{\rm{eff}}=707$ Hz,
while simultaneously being modulated at a much lower frequency of
about 11 Hz.

In order to visualize how the spin and motional dynamics become
entangled over a time long compared to the Rabi period, we show
snapshots of the density of each state in Figure 3.  Three different
sets of snapshots are shown, corresponding to the three circled
numbers in Figure 2. A full Rabi cycle is shown for each set. The
first set begins at $t=0$ with all of the atoms in the $|1\rangle$
internal state and in the mean-field ground state of the trap
$|\phi_0\rangle$. During this first Rabi cycle, the shape of the
density profile for each internal state does not change much---only
the height changes. That is, the motional state remains the ground
state while population cycles rapidly between internal states, as
discussed below Eq.~(\ref{twostatelab}).

\begin{figure}
  \centerline{\epsfig{file=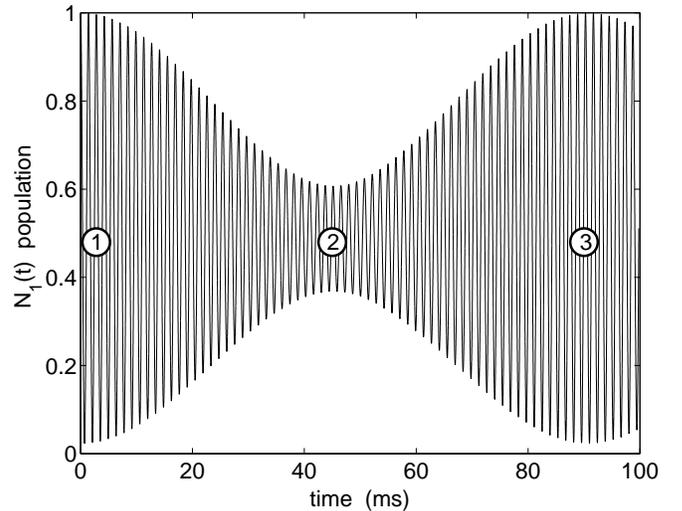,height=2.7in}}
\vspace{0.05cm}
\caption{This plot shows an example of the modulation of the Rabi
oscillations. The fractional population in state $|1\rangle$ is
plotted as a function of time, obtained from a numerical solution of
the one dimensional version of Eq.~(\ref{main1}). The values of the
various parameters are given in the text. In Figure 3, the densities
for both states are shown for three different Rabi cycles designated
by the circled numbers in this plot.}
\end{figure}
\begin{figure}
  \centerline{\epsfig{file=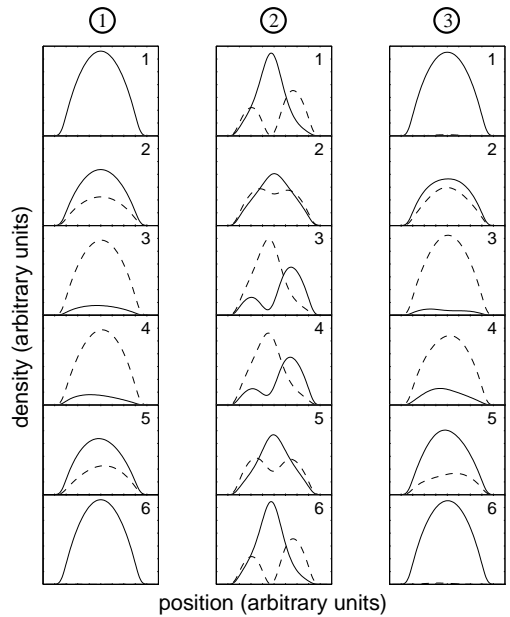,height=3.2in}}
\vspace{0.05cm}
\caption{This plot shows snapshots of the density of each state for
three different Rabi cycles corresponding to the three circled numbers
in Figure 2. The solid line is the density of the $|1\rangle$ state,
while the dashed line is that of the $|2\rangle$ state. Each snapshot
within a set is numbered in sequential order.  The first set starts at
$t = 0$ ms, and runs for a full Rabi cycle 1.41 ms. The second and
third sets begin at $t = 45.2$ ms and $t = 90.3$ ms, respectively.
The time increment between snapshots is $\Delta t= 0.28$ ms for all
three sets.}
\end{figure}

\begin{figure}
  \centerline{\epsfig{file=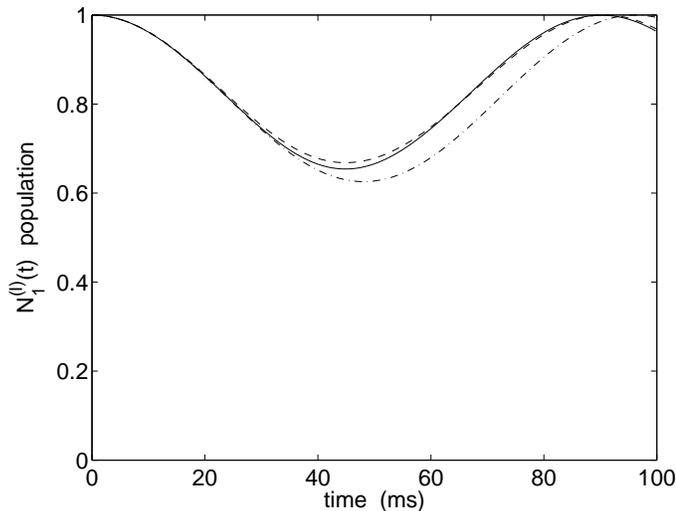,height=2.7in}}
\vspace{0.05cm}
\caption{The fractional population of the $|1\rangle$ state in the
rotating frame is shown. The solid line is the solution given by the
CGSD model, while the dot-dashed line corresponds to the solution of
the two-mode model. If the two-mode model is extended to include
coupling to the first even-parity excited mode, then we get better
agreement to the CGSD model, as shown by the dashed line.}
\end{figure}
\begin{figure}
  \centerline{\epsfig{file=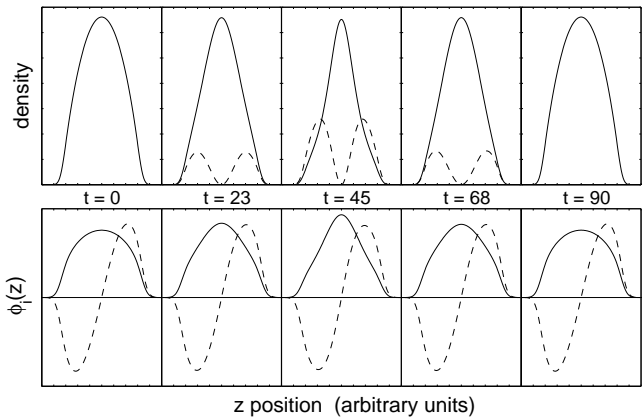,height=2.2in}}
\vspace{0.05cm}
\caption{The top strip of this plot shows snapshots of the density of
each state corresponding to the solution of the CGSD model given by
the solid line in Figure 4. The bottom strip shows the corresponding
two self-consistent eigenmodes given by the solution of
Eq.~(\ref{selfcons}). The times of each snapshot are shown in the
region between the two strips, in units of milliseconds. The solid
line corresponds to the density of the $|1\rangle$ state, while the
dashed is that of the $|2\rangle$ state.}
\end{figure}
The second set of snapshots in Figure 3 is taken at around $t=45$ ms,
which is halfway through the modulation. The density profiles for each
spin state cycle rapidly between a single-peaked and a double-peaked
structure.  For example, in the first snapshot, the $|1\rangle$ state
is in the single-peaked structure, while the $|2\rangle$ state is in
the double-peaked structure, but halfway through the Rabi cycle the
situation is reversed, as shown in the third and fourth
snapshots. Finally, at about $t=90$ ms when the amplitude of the Rabi
oscillations has revived, the third set shows that the motional and
spin degrees of freedom appear to be decoupled again, with the density
profile of each spin state appearing as it did during the first Rabi
cycle.

As outlined in Section II, this peculiar behavior is most easily
understood by going to the rotating frame. In Figure 4, we plot the
fractional population in the $|1\rangle$ state in the rotating frame
$N_1^{(I)}(t) = \int |\langle z | \langle 1|\psi^{(I)}(t)\rangle|^2
dz$.  The solid line corresponds to the CGSD model presented in
Section IID. In the rotating frame, population is slowly transferred
out of the $|1\rangle$ state due to the coupling from
${\hat{H}}_z'\otimes {\hat{\sigma}}_x$ in Eq.~(\ref{HofI2}).

In the rotating frame, the system is being excited out of the ground
state $|\phi_0\rangle$ due to the dipole coupling $H_z'$.  This can be
seen in the top strip of snapshots in Figure 5, where the density of
each spin state in the rotating frame is shown, corresponding to the
solid line in Figure 4.  Initially, all of the atoms are in the
$|1\rangle$ internal state and the mean-field ground state of the trap
$|\phi_0\rangle$.  Due to the dipole coupling, population is
transferred out of the ground state.

The strongest coupling is between the ground $|\phi_0\rangle$ and the
first excited $|\phi_1\rangle$ modes. These eigenmodes are shown in
the bottom strip of Figure 5.  They evolve slowly in time as the
coefficients $c_0(t)$ and $d_1(t)$ change.  For example, initially the
ground state is just the Thomas-Fermi-like ground state, since all of
the population is in that state.  However, at $t=45$ ms, about
one-third of the population is in the first excited mode, which
pinches the ground state due to the mean-field interaction term in
Eq.~(\ref{HSA2}). That is why the self-consistent ground state at
$t=45$ ms is narrower than at $t=0$.

It is clear from Figure 4 that the low-frequency modulation of the
rapid Rabi oscillations in the lab frame is just the frequency of
oscillation in the rotating frame between $|\phi_0\rangle|1\rangle$
and $|\phi_1\rangle |2\rangle$. This is reflected in the two-mode
solution given by Eq.~(\ref{twostatelab}), which also helps explain
the peculiar behavior of the densities shown in Figure 3.  In the lab
frame the system is cycling rapidly between the two modes shown in
Figure 5. The initial values of the energies are $\epsilon_0=13.6 \,
\hbar \omega_z$ and $\epsilon_1=13.7 \, \hbar \omega_z$, which makes
$\Delta_{01}=0.1 \, \hbar \omega_z$. This small energy splitting is
due to the effect of the mean field, since in the limit
$N\rightarrow1$ these energies move apart by a factor of ten, which
greatly reduces the coupling between the modes and thus greatly
reduces the modulation effect.

If we make the two-mode ansatz and solve Eq.~(\ref{twostateevolve}),
we get the dot-dashed line in Figure 4. The discrepancy from the solid
line arises due to a weak coupling between the first $|\phi_1 \rangle$
and second $|\phi_2\rangle$ excited modes.  If we extend our two-state
model to include this third mode, we get the dashed line in Figure 4,
which nearly sits on top of the solid line. In this case, the second
excited mode $|\phi_2\rangle$ gains less than $5 \%$ of the total
population.

\begin{figure}
  \centerline{\epsfig{file=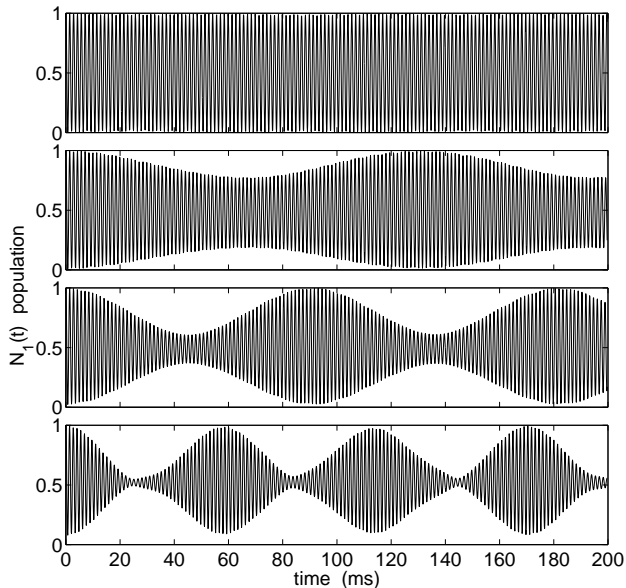,height=3.1in}}
\vspace{0.05cm}
\caption{This plot shows the fractional population in the $|1\rangle$
state for four different values of the detuning, obtained from a
numerical solution of Eq~(\ref{main1}). Starting from the top, the
detuning is $\delta = 0$, $\delta = 50$ Hz, $\delta = 100$ Hz, and
$\delta = 200$ Hz. The values of the other parameters are given in the
text.}
\end{figure}
\begin{figure}
  \centerline{\epsfig{file=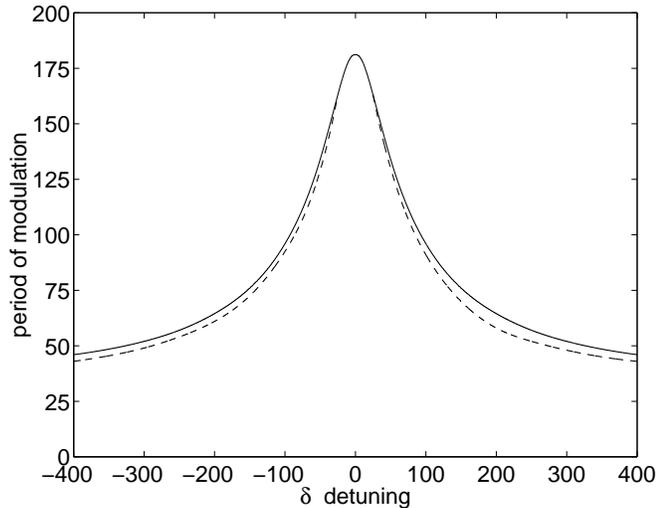,height=2.7in}}
\vspace{0.05cm}
\caption{This plot shows the period of modulation as a function of
detuning $\delta$. The dashed line corresponds to the numerical
solution of the one-dimensional version of Eq.~(\ref{main1}), while
the solid line was obtained from a numerical solution of the two-mode
model Eq.~(\ref{twostateevolve}). The Rabi frequency was $\Omega =
700$ Hz. }
\end{figure}

\subsection{Dependence on detuning}
In Figure 6, we show how the behavior of the system depends on the
detuning $\delta$.  The Rabi frequency $\Omega=700$ Hz was held fixed
for each plot while the detuning was varied from zero at the top
$\delta = 0$ to $\delta = 200$ Hz in the bottom plot. As predicted by
the coupling parameter $\beta =z_0 \, \delta \, \Omega /
\Omega_{\rm{eff}}^2$ in the CGSD model, no coupling between motional
states occurs if $\delta = 0$, and thus the Rabi oscillations
experience no modulation.  As $\delta$ is increased the motional-state
coupling becomes stronger and we expect the modulation frequency to
increase. The amplitude of modulation also increases as the detuning
is increased.

We show the dependence of the period of modulation on detuning more
explicitly in Figure 7. The dashed line is the numerical solution of
the full problem given by Eq.~(\ref{main1}), while the solid line is
the numerical solution of the two-mode model given by
Eq.~(\ref{twostateevolve}).

\subsection{Dependence on trap displacement}
In Figure 8, we show how the behavior of the system depends on the
trap displacement $z_0$. The Rabi frequency $\Omega=700$ Hz and the
detuning $\delta = 100$ Hz were held fixed, while the trap
displacement was varied from zero $z_0=0$ in the top plot to $z_0 = 1
\, \mu$m in the bottom plot. Again, the coupling parameter $\beta$
predicts no modulation if $z_0=0$. As $z_0$ is increased, the
frequency of modulation increases as the system is driven harder.
However, for the large separation in the bottom plot, the modulation
becomes highly irregular and the two-mode model most certainly breaks
down. This behavior may be chaotic and warrants further investigation.
\begin{figure}
  \centerline{\epsfig{file=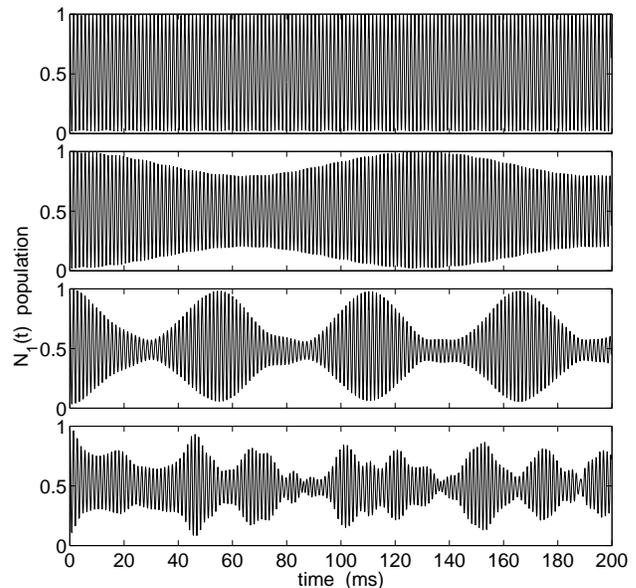,height=3.1in}}
\vspace{0.05cm}
\caption{This plot shows the fractional population in the $|1\rangle$
state for four different values of the trap displacement $z_0$,
obtained from a numerical solution of Eq~(\ref{main1}). Starting from
the top, the displacement is $z_0=0$, $z_0 = 0.1 \, \mu $m, $z_0 = 0.4
\, \mu $m, and $z_0 = 1.0 \, \mu $m. The values of the other
parameters are given in the text.}
\end{figure}

\section{Conclusions}\label{conclusion}

The gross features predicted by our model, such as double-peaked
structure in the density distribution, and the presence of collapses
and revivals in the relative population dynamics, are supported by
experimental observation~\cite{exp5}.  Experiment-theory agreement on
finer points is only fair.  The theory tends to underestimate the
contrast ratio of the collapses, for instance.  Moreover, to match the
detuning trends shown in Fig. 6 and Fig. 7 one needs to add by hand an
unexplained overall detuning offset. This is most likely due to there
actually being a spatial dependence of the bare Rabi frequency due to
the influence of gravity on the untrapped intermediate state of the
two-photon transition. To model the experimental situation in more
detail one would have to include this effect as well as inelastic loss
processes and finite-temperature effects neglected here. It may be
also that treating the TOP trap potential as purely static may be an
oversimplication.

In this paper we have demonstrated the possibility for quantum state
engineering of topological excitations through the interplay between
the internal and spatial degrees of freedom in a Bose condensed
gas. Due to the symmetry of the system we have analyzed, the
excitation in our case was the odd-parity dipole mode. The intriguing
possibility of exciting modes with alternative symmetries, such as a
vortex mode~\cite{Marzlin1,Walls,Zoller,Bagnato}, would require a
different trap geometry, but is a straight-forward extension of the
analysis presented here. Although we have focussed in this work on a
particular parameter regime, the system is a rich one for study and
exhibits complex and perhaps chaotic dynamics under strong excitation
conditions.

\section{Acknowledgments}

We would like to thank Howard Carmichael for highlighting the
analogies between this system and the bichromatically driven two-level
atoms~\cite{howard}, and also Allan Griffin and Eugene Zaremba for
insightful discussions. Finally, we would like to thank David Hall,
Mike Matthews, and Paul Haljan for working in parallel on the
experimental side of this project and for sharing the results of their
observations in the laboratory~\cite{exp5}. This work was supported by
the National Science Foundation. E.C. would also like to thank the
Office of Naval Research and the National Institute for Standards and
Technology for funding support.

\bibliographystyle{prsty} \bibliography{colrevbib}

\begin{table}[t]
\vspace{1cm}
\begin{center}
\caption{This is a table showing the values used for the various
physical parameters appearing in our calculations. The scattering
lengths are taken from \protect\cite{exp2}.}
\begin{tabular}{llcll}
  $N$ & $8 \times 10^5$ & \vline & $\nu_z$ & $65 \, {\rm{Hz}}$ \\
$a_{21}$ & $5.5(3) \,{\rm{nm}}$ & \vline & $\nu_{\rho}$ & $24 \,
{\rm{Hz}}$ \\ $a_{22}$ & $0.97 \, a_{21}$ & \vline & $z_{\rm{sho}}$ &
$1.3 \, \mu$m \\ $a_{11}$ & $1.03 \, a_{21}$ & \vline & $z_0$ & $0.2
\, \mu\rm{m}$ \\ \end{tabular} \end{center} \label{table1}
\end{table}

\end{document}